# WIP: Software Engineering Competencies in the Age of AI

Lynn Vonderhaar
*Department of Electrical Engineering and Computer Science*
*Embry-Riddle Aeronautical University*
Daytona Beach, USA
vonderhl@my.erau.edu

Massood Towhidnejad
*Department of Electrical Engineering and Computer Science*
*Embry-Riddle Aeronautical University*
Daytona Beach, USA
towhid@erau.edu

Omar Ochoa
*Department of Electrical Engineering and Computer Science*
*Embry-Riddle Aeronautical University*
Daytona Beach, USA
ochoao@erau.edu

***Abstract*— A university education is meant to fully prepare graduates to enter and succeed in the workforce in their field. Accreditation bodies and professional organizations help to achieve this by publishing curriculum content, skills and competency models to outline the basics that universities must teach within a degree program. Although these curriculum guidelines and competency models are periodically updated to align university curricula with current industry needs, the needs of industry evolve rapidly and are often not properly reflected within the model, a situation which is occurring now with the rise of Artificial Intelligence (AI) use in industry. This paper reviews literature describing the AI needs within industry to motivate two new AI competencies, AI Literacy and AI development, within the Software Engineering Competency Model (SWECOM). This work then analyzes the industry needs described in the literature to propose the necessary skill areas for these new competencies.**

***Keywords—SWECOM, software engineering, competencies, artificial intelligence***

## I. Introduction

One of the primary missions of a university is to serve as an institution of higher learning by providing high-quality education that fosters critical thinking, creativity, and intellectual growth. In this context, software engineering programs are expected to prepare graduates to design, develop, and maintain high-quality software systems that meet real-world needs. These real-world needs are defined by university accreditation bodies and professional organizations that publish required curriculum content, skills and competencies for graduates. The SE2004 (Software Engineering 2004) [1] and SE2014 (Software Engineering 2014) [2], provide comprehensive recommendations for undergraduate software engineering degree programs. On the other hand, the Software Engineering Competency Model (SWECOM) [3], and Computing Curricula 2020 (CC2020) [4] outline the necessary competencies for graduates of a software engineering program. Although these guidelines and recommendations are periodically updated, the rapid pace of technological advancement in the software industry and the increasing complexity of software systems present significant challenges for universities and organizations in keeping curricula current and relevant. This phenomenon is relevant today as graduates enter the workforce in the Golden Age of Artificial Intelligence (AI), where the competencies that they gained during their education may not reflect the AI skills necessary to succeed in industry.

This shortcoming is highlighted by the fact that the Software Engineering Body of Knowledge (SWEBOK) recognizes the importance of AI. The SWEBOK is a guide published by the IEEE which defines the knowledge, best practices, and standards within the software engineering profession. SWEBOK is frequently updated and in the latest version of this document (SWEBOK V4) [5], the area of Artificial Intelligence has been identified as one of the foundational knowledge areas, while SWEBOK V3 [6], which was published in 2014, did not mention AI. This signals the importance of AI in the current state of the software development industry. Unfortunately, AI is less prevalent in the software engineering curriculum guidelines. The latest curriculum guideline for an undergraduate software engineering degree, as defined by SE 2014, does not mention AI as an area of importance. The same is true for the SWECOM which establishes the necessary competency for software engineers. Currently, we are investigating the effect of AI on undergraduate software engineering curriculum and the expected competencies for recent graduates of these programs as it relates to the software industry expectations. We believe the existence of this gap will make the graduates of the software engineering programs ill-prepared for their first career. Our work has two tracks, one concentrating on the identification of the curriculum gap, and another identifies the existing gap in the competency.

This Work-In-Progress (WIP) paper concentrates on the competency gap, by examining the literature and identifying the AI competencies within undergraduate software engineering programs. This study seeks to highlight the gap between academic curricula and industry expectations for entry-level software engineers entering the workforce in this Golden Age of AI, and to set the stage for what an AI competency and the corresponding skill areas might look like within SWECOM based on findings within the literature.

## II. Background

The SWECOM is a conglomeration of several other existing competency models including:

- The engineering competency model,
- Information technology,
- Systems engineering,
- Software assurance,
- SWEBOK,

- ISO/IEEE Standard 12207 on Systems and software engineering – software life cycle processes [7],
- Graduate Software Engineering Curriculum Guidelines (GswE2009) [8], and
- Undergraduate Software Engineering Curriculum Guidelines (SE2004).

Importantly, SWECOM separates competency and skill acquisition from specific degree requirements and a person's number of years of experience. SWECOM focuses on 13 areas of technical skills [3]. Five skill areas correlate to specific phases of the Software Development Life Cycle (SDLC), while the other eight areas crosscut the phases within the SDLC. The skill areas within SWECOM are shown in Table I.

TABLE I. SWECOM SKILL AREAS

| Software Life Cycle Skill Areas | |
|---|---|
| Software Requirement | Requirement Elicitation |
| | Requirement Analysis |
| | Requirement Specification |
| | Requirement Validation and Verification |
| | Requirement Product and Process Management |
| Software Design | Design Fundamentals |
| | Design Strategies and Methods |
| | Architecture Design |
| | Design Analysis and Evaluation |
| Software Construction | Construction Planning |
| | Construction Management |
| | Detailed Design and Coding |
| | Debugging and Testing |
| | Integration and Collaboration |
| Software Testing | Test Plan |
| | Test Infrastructure |
| | Test Techniques |
| | Test Measurement and Defect Tracking |
| Software Sustainment | Software Transition |
| | Software Support |
| | Software Maintenance |
| Software Engineering Cross Cutting Skill Areas | |
| Software Process Life Cycle | Requirements |
| | Design |
| | Construction |
| | Testing |
| | Process |
| | Quality |
| Software Systems Engineering | Life Cycle Concept Definition |
| | Requirements Engineering |
| | Design |
| | Requirements Allocation |
| | Component Engineering |
| | Integration and Verification |
| | Validation and Deployment |
| | Sustainment Planning |
| Software Quality | Software Quality Management (SQM) |
| | Reviews |
| | Audits |
| | Statistical Control |
| Software Security | Requirements |
| | Design |
| | Construction |
| | Testing |
| | Process |
| | Quality |
| Software Safety | Requirements |
| | Design |
| | Construction |
| | Testing |
| | Process |
| | Quality |
| Software Configuration Management | Plan Configuration Management |
| | Conduct Configuration Management |
| | Manage Software Releases |
| Software Measurement | Plan Measurement Process |
| | Perform Measurement Process |
| Human Computer Interaction | Requirements |
| | Interaction Style Design |
| | Visual Design |
| | Usability Testing and Evaluation |
| | Accessibility |

SWECOM also establishes levels of competencies: technician, entry-level practitioner, practitioner, technical leader, and senior software engineer [3]. This work focuses on competencies for an entry-level practitioner, which is the level at which most graduates will enter industry upon receiving their bachelor's degree.

Our previous work reviews the SWECOM guidelines and surveys four undergraduate software engineering programs in their preparation of their graduates based on SWECOM [9]. The purpose of our previous work was to both review the competencies within SWECOM and to understand how well accredited software engineering programs in United States universities are aligning to SWECOM and preparing their students for industry. The work presented here will build on this previous work to establish AI as a necessary competency for software engineering students. This work, therefore, sets up future work in analyzing how accredited universities prepare their graduates for AI competency within industry.

## III. LITERATURE REVIEW

To support the establishment of AI as a necessary competency within SWECOM, this work presents a literature review of AI requirements within industry. This includes both the use of AI to aid in entry-level jobs, i.e., AI-enabled software engineering, as well as the engineering of AI systems.

### A. Search Method

Although this work is not meant to be a systematic literature review, this subsection provides a search method and inclusion criteria for the reader to better understand the provided results. The search phrases were meant to provide a broad scope of literature which could be pruned to the inclusion criteria. Table II shows the search terms and the resulting references. These references were found using the respective search terms on Google Scholar.

Table II. Search terms and references for this literature review.

| **Search Terms** | **References** |
|---|---|
| educating software engineers in the age of AI | [10, 11, 12, 13, 14] |
| software engineering graduates AND industry AND AI | [15, 16] |
| AI job skills revolution AND software engineering | [17, 18] |
| AI job AND competence AND software engineering | [19, 20] |
| AI AND machine learning AND job advertisement | [21, 22] |

After gathering sources from the search terms shown in Table II, the base of sources was pruned using the following inclusion criteria:

- The paper must be written in English.
- The paper must be published in a peer reviewed conference or journal.
- The paper must relate to AI or Machine Learning (ML) skills sought after in industry.

### *B. AI-Enabled Software Engineering*

Much of literature focuses on how AI-based tools are used in developing other software, whether it be other traditional software, or other AI-based software. Kam, et al. note the increasing use of AI-based tools, e.g., Large Language Models (LLMs) to improve productivity in industry jobs [13]. However, the authors also state that developers who attempt to use AI-based tools without proper preparation and instruction may actually take longer to perform their task. The paper analyzes the use of generative AI by employees who successfully use the tools to increase their productivity. The authors seek to answer the questions of what tasks these employees use generative AI for and what skills and knowledge they have about the tools that allow them to use the tools so effectively. These insights can be applied to preparing graduates with skills now necessary when entering industry as a software developer, even if they are not developing AI-based systems.

Several other authors specifically note the importance of incorporating AI literacy into university education [10, 11]. Marlowe, et al. perform a literature review to show how AI-based tools are revolutionizing the field of software engineering [10]. The authors outline uses for AI within industry including but not limited to writing code, analyzing requirements, reducing requirement ambiguities, recommending design patterns, and efficiently reviewing bug reports. This builds to the argument that software engineering education needs to adapt to the Age of AI where the minute details of syntax and coding may be less relevant and understanding and interacting with AI-based tools to produce software is crucial.

Kumar Sah, et al. also identify the need for AI literacy education, but also note that there are several challenges that universities are currently facing to making this a reality including educators that are inexperienced in AI and resource limitations [11]. The authors note that there are multiple ways to incorporate AI into software engineering curricula including dedicated classes, as well as new AI modules built into existing classes.

### *C. Engineering AI Systems*

Other literature focuses on the competencies necessary for developing AI-based systems. For instance, Modran, et al. discuss the reskilling initiative of prepping students for how AI will change jobs [18]. They include the following topics as necessary for succeeding in today's workforce: choosing the best model architecture, exploratory data analysis, data cleaning techniques, model training, and model testing.

Similarly, Shaji George and Hovan George list several new roles emerging within industry that relate to the incorporation of AI within existing pipelines [17]. They also analyze new skills needed when entering the workforce today, including new development libraries, statistical knowledge, and data visualization skills. The authors also stress the increased importance of communication skills as many business stakeholders may struggle to keep up with the rapidly evolving trends, making it crucial that developers be able to explain complex concepts to other stakeholders. The authors, however, focus on how companies and organizations can cope in the Golden Age of AI, but do not discuss the growing need for universities to adapt.

Other authors focus on specific classes that they have run to improve students' foundational knowledge of AI [15, 16]. Palomba, et al. provide an experience report based on offering a Software Engineering for AI (SE4AI) course that addressed differences between engineering traditional software systems versus AI-based systems [12]. The authors note that not only is it important to address the developmental differences in AI-based systems, but also the new quality attributes that come with developing AI, e.g., robustness and ethics, which are not concerns in traditional software systems. The experience report includes insights on specific challenges that students faced over four iterations of this course and stresses the importance of offering hands-on experience to strengthen students' practical knowledge of development. The authors also note that AI technologies evolve rapidly, making it important to also change the course content often to remain relevant. Similarly, Chenoweth and Linos describe a new undergraduate class that combines agile software engineering with machine learning [16]. The authors note that using agile development methods when developing ML-based systems can be particularly challenging because the ML development process often does not yield a complete deliverable until the project nears conclusion. The authors' goal was to address these challenges within their class and teach students to incorporate agile practices into ML development.

More broadly than specific classes, Bublin, et al. offer suggestions to adapting SE curriculums specifically for developing AI systems [14]. The authors note how important it is to change SE curriculum for AI systems because of the data-centricity of AI systems, making many traditional development approaches not applicable to these systems. Their guidelines are based in literature and experience in industry. The authors lay out the differences between traditional software and AI systems within several phases of development and outline the main challenges for engineering AI systems.

Other literature builds skill profiles from job postings [18, 19, 20]. Some authors research the necessary skills for the AI job market by analyzing job postings [18, 19]. Jia, et al. build a knowledge graph of necessary skills for succeeding in an AI job [21]. The knowledge graph's structure allows the authors to draw connections between types of jobs and skills that they require. For example, a deep learning engineer should have skills in Tensorflow, convolutional neural networks, recurrent neural networks, and CUDA. Meanwhile a computer vision engineer would also need skills in OpenCV. It is important to note that this paper was published in 2018 and therefore does not reflect all current competencies. Meanwhile Verma, et al. analyze the differences in skills required for jobs titled with "AI" versus those titled with "ML" [22]. It therefore reasons that universities that want to provide their students with opportunities in these fields should embed such skills within their curricula. Another paper by Brauner, et al. specifically notes a lack of AI competencies within industry, specifically industry within the United Kingdom (UK) [19]. The authors note that industry is not properly defining the desired competencies and that universities are not changing their curricula fast enough to meet industry's changing needs. The paper outlines specific competencies for data science, AI software development, AI product development and management, AI client servicing, and AI research.

## IV. Related Work

There are several papers that outline changes to SE curriculum or even offer specific class solutions to better prepare graduates for industry in the Golden Age of AI [14, 12, 10, 16]. Bublin, et al. discuss ways in which the SE curriculum could be adapted to account for AI-based systems [14]. The authors outline potential changes to curriculum based on findings in literature and industry experience. Their literature search showed several key challenges that developers face when engineering AI-based systems, which could guide potential curricular changes. While the authors offer suggestions for changing university curriculum to better prepare graduates for jobs in AI engineering, they do not discuss the importance of also changing the regulatory requirements that guide curriculum development and accreditation, which this work does. Meanwhile Palomba, et al. offer insights into teaching SE4AI from a specific course taught on the subject, which again does not address accreditation concerns, or even larger curricular changes [12].

Marlowe, et al. identify AI literacy as a core topic in software engineering education today [10]. Their literature review builds the argument that generative AI and other AI tools are consistently being used in software development in industry, making it crucial that students graduating today be adept at using them. The authors also stress the importance of graduates understanding the underlying concepts of AI and machine learning so that they can interpret and combat common AI challenges, e.g., hallucinations and algorithmic bias. Throughout their paper, the authors note areas of SWEBOK, but do not mention how AI might be included in it or other guidelines.

Brauner, et al. do specifically define competencies for five types of AI jobs within the UK [19]. Specifically, they define competencies for data science, AI software development, AI product development and management, AI client servicing, and AI research. However, the authors are only defining competencies for specific jobs based on job postings. They are not integrating these competencies into curricula of a broader field, e.g., software engineering, which this work does. The goal of our recent project is based on the latest version of SWEBOK, where specifically, AI is identified as one of the key fundamentals of computing that is needed for software engineers. To conclude, we believe both the software engineering curriculum guideline and the software engineering competency need to be updated to reflect the latest changes in the industry in the last 12 years, which has not been addressed in previous work.

## V. Discussion

The purpose of this WIP paper is to establish the need for AI as a competency within SWECOM and propose the skill areas within that competency. As the literature review shows, there are two main skill areas required within the AI competency for SWECOM: AI literacy and AI development. Based on specific skills mentioned within the reviewed literature, Table III shows the proposed skill areas for an AI competency in SWECOM. The proposed skill areas are based on the skills discussed directly in the reviewed literature.

Table III. Proposed SWECOM skill areas for AI

| AI Literacy | |
|---|---|
| Generative AI | Prompt engineering |
| | Assessing the quality of outputs |
| | Ethical and responsible use |
| **AI Development** | |
| Foundational Knowledge | Statistical literacy |
| | Python |
| | Machine learning Python libraries |
| Data Engineering | Exploratory Data Analysis |
| | Data cleaning and preprocessing |
| | Data visualization |
| Classical AI Approaches | Regression methods |
| | Classification methods |
| | Ensemble methods |
| Deep Learning Approaches | Neural Networks |
| | Large Language Models |
| | Transfer learning |
| ML Engineering | Requirements Engineering for AI-based systems |
| | Architectural design |
| Quality Assurance | Testing |
| | Performance metrics |
| | Ethical and responsible use |
| | Data drift |

Generally, literature discussing the use of AI-enabled software engineering stresses the need for AI literacy in the use of generative AI and more specifically, the use of LLMs. The literature notes the need for competency in effectively prompting the model for the desired results, but also in assessing the quality, bias, and ethics of the generated result.

Meanwhile the AI development skill area is much more involved. This skill area requires not only foundational knowledge that ML is based on, but also several topics within data engineering, knowledge of multiple model types, changes to requirements engineering and architectural design, and finally methods of quality assurance for AI-based systems.

## VI. Limitations and Future Work

This WIP paper proposes adding additional AI content to the curriculum and AI competency to SWECOM and offers suggested skill areas to accomplish this. In another word, adding two additional Cross Cutting Skill Areas. These changes would align the curriculum guidelines and SWECOM with the newest version of SWEBOK. The future work for this includes three main tracks: one concentrates on the identification of the curriculum gap, another identifies the existing gap in the competency, and a third integrates these new competencies into the existing SWECOM.

However, there are also some important concerns that future work should address while investigating these curricular guidelines. First, several authors note in the literature that AI technology and relevant knowledge changes rapidly [12, 17, 20, 15]. It therefore may be particularly difficult for the curriculum and competency committees to keep up with the changing demands of industry. Future work will analyze this challenge and provide a plan to address it.

It is also important to consider challenges in fitting additional content and competency into already full curricula. Whether this involves dedicated classes or modules within existing courses, future work needs to analyze what other content and competencies might be weakened by adding these new topics. It is also critical to have a plan to maintain modern curricula based on the rapidly evolving AI field.

Finally, some literature has noted that some universities may struggle to find educators that are experienced enough in these new AI skill areas to teach them to their students [11]. It is therefore crucial that future work explore these challenges and propose solutions for universities before adapting their curricula.

## VII. Conclusion

The purpose of this work was to propose the addition of an AI competency to SWECOM and suggest skill areas for students to achieve this new competency. The motivation behind the new competency is grounded in a literature review, which finds that skills in both AI literacy and AI development are now highly coveted in industry. It is therefore crucial that universities properly prepare their graduates in this area to provide their graduates with those opportunities within industry. Based on the conducted literature review, this work developed potential skill areas to form the new AI competency within SWECOM. Future work for this paper includes exploring a plan for maintaining up-to-date skill areas and analyzing necessary changes to existing software engineering curricula to make room for additional competency.